\documentclass[conference]{IEEEtran}
\IEEEoverridecommandlockouts
\usepackage{cite}
\usepackage{amsmath,amssymb,amsfonts}
\usepackage{algorithmic}
\usepackage{graphicx}
\usepackage{textcomp}
\usepackage{xcolor}
\usepackage[a4paper, total={184mm,240mm}]{geometry}

\def\BibTeX{{\rm B\kern-.05em{\sc i\kern-.025em b}\kern-.08em
    T\kern-.1667em\lower.7ex\hbox{E}\kern-.125emX}}

\usepackage{comment}
\usepackage{romannum}

\usepackage{tikz,xcolor,hyperref}

\definecolor{lime}{HTML}{A6CE39}
\DeclareRobustCommand{\orcidicon}{%
	\begin{tikzpicture}
		\draw[lime, fill=lime] (0,0) 
		circle [radius=0.16] 
		node[white] {{\fontfamily{qag}\selectfont \tiny ID}};
		\draw[white, fill=white] (-0.0625,0.095) 
		circle [radius=0.007];
	\end{tikzpicture}
	\hspace{-2mm}
}
\newcommand{\orcidKR}{\href{https://orcid.org/0000-0002-4555-3931}{\orcidicon}}
\newcommand{\orcidPM}{\href{https://orcid.org/0000-0003-3653-6221}{\orcidicon}}

\usepackage[utf8]{inputenc}
\begin{document}

\title{Modeling and Exploration of Gain Competition Attacks in Optical Network-on-Chip Architectures
}


\author{
\IEEEauthorblockN{Khushboo Rani~\orcidKR, Hansika Weerasena, Stephen A. Butler, Subodha Charles, and Prabhat Mishra~\orcidPM
}\\
\vspace{-4mm}
\IEEEauthorblockA{\textit{CISE, University of Florida, Gainesville, Florida, USA}}
}

\maketitle

\begin{abstract}
Network-on-Chip (NoC) enables energy-efficient communication between numerous components in System-on-Chip architectures. The optical NoC is widely considered a key technology to overcome the bandwidth and energy limitations of traditional electrical on-chip interconnects. While optical NoC can offer high performance, they come with inherent security vulnerabilities due to the nature of optical interconnects. 

In this paper, we investigate the gain competition attack in optical NoCs, which can be initiated by an attacker injecting a high-power signal to the optical waveguide, robbing the legitimate signals of amplification. 
To the best of our knowledge, our proposed approach is the first attempt to investigate gain competition attacks as a security threat in optical NoCs. We model the attack and analyze its effects on optical NoC performance. We also propose potential attack detection techniques and countermeasures to mitigate the attack.
Our experimental evaluation using different NoC topologies and diverse traffic patterns demonstrates  the  effectiveness  of  our  modeling  and exploration of gain competition attacks in optical NoC architectures.

\end{abstract}

\section{Introduction} \label{section:introduction}

With the increasing levels of instruction, thread, and data-level parallelism in CPUs, communication between diverse components in System-on-Chip (SoC) architectures has become a major bottleneck from both performance and energy perspectives. There has been a shift toward utilizing Network-on-Chip (NoC) based communication in designing SoCs. Unlike bus-based architectures, NoC offers a scalable and efficient solution~\cite{dally2001route}. 
 In the early days, NoC design relied on electrical interconnects to communicate between SoC components. 
However, electrical interconnects face a serious challenge in keeping up with the rate at which data must be transferred in the presence of increasing parallelism and decreasing chip area. Drawing from the domain of telecommunications, optical/photonic interconnects provide higher bandwidth, better latency, reliability, and less energy usage compared to their electrical counterpart~\cite{pasricha2020survey}. Several prior studies explored optical interconnects as a replacement for electrical on-chip interconnects~\cite{xie2010crosstalk}~\cite{pasricha2020survey}. 
While Optical Network-on-Chip (ONoC) can offer high performance, it comes with inherent security vulnerabilities due to the nature of optical interconnects. 
SoC manufacturers use third-party Intellectual Property (IP) blocks in their designs to reduce cost and meet time-to-market constraints. The long supply chains coupled with the utilization of potentially untrusted IPs increases the likelihood of different types of security threats, such as hardware Trojans, for eavesdropping or degrading Quality-of-Service (QoS)~\cite{charles2020real}~\cite{chittamuru2020exploiting}. 

In the domain of optical communications, Erbium-Doped Fiber Amplifiers (EDFAs) or Semiconductor Optical Amplifiers (SOA)~\cite{connelly2007semiconductor} are used to periodically strengthen an optical signal to counteract attenuation and signal drops~\cite{chittamuru2018soteria}. Similarly, optical interconnects on the chip level can also experience attenuation and amplitude drops. Grani and Bartolini~\cite{10.1145/2602155} quantified these losses that can happen over the optical waveguide and at different points in the network. Such drops include a 0.02dB drop at a key component in the ONoC, called the Microring Resonator (MR). MRs are critical to filtering wavelengths in the photonic modulators, which are responsible for electrical to optical conversions~\cite{pasricha2020survey}. Amplitude drops can be more severe in Dense Wavelength-Division Multiplexed (DWDM) networks when there are potentially $64$ or $128$ MRs (one for every wavelength) at one point in the network, causing a drop of 1.28 ($0.02 \times64$) or 2.56 ($0.02 \times128$) dB per node that the light passes through. Other drops include a waveguide propagation loss of $1$ dB/cm and a $0.7$ dB loss per waveguide cross~\cite{10.1145/2602155}. These losses introduce the need for {\it optical amplifiers} at the chip scale so that signals can remain at a reasonable power level for photodetectors.

    \begin{figure}
        \centering
        \includegraphics[scale=0.6]{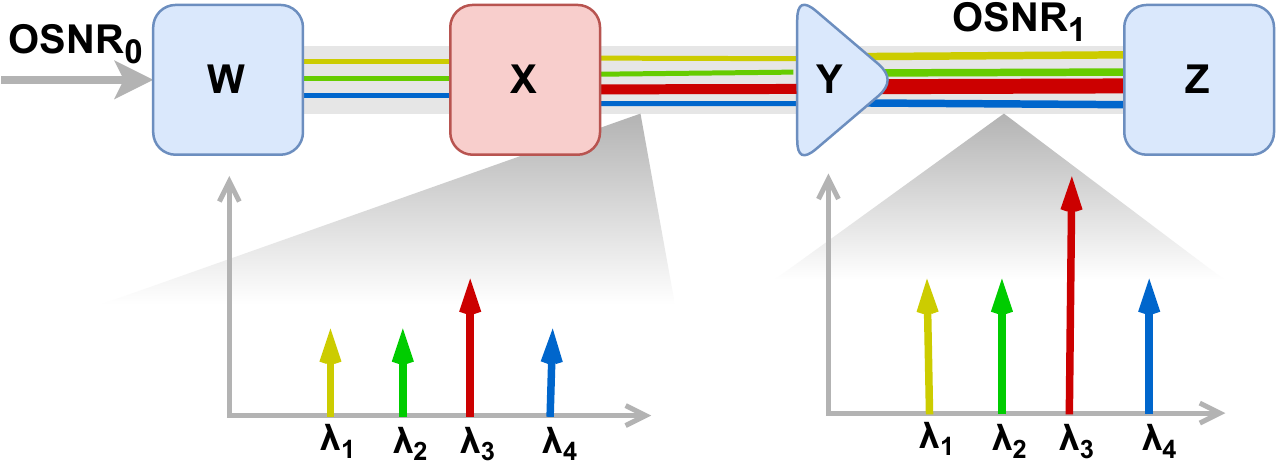}
        \vspace{-2mm}
        \caption{Gain Competition Attack: The malicious wavelength ($\lambda_3$) gets higher amplification, which steal the power needed for amplification of the legitimate wavelengths ($\lambda_1,\lambda_2$, and $\lambda_4$).}
        \label{fig:edfaBig}
        \vspace{-0.2in}
    \end{figure}


\subsection{Threat Model}

A gain competition attack occurs when a malicious party injects a powerful optical signal into the network, and on amplification, the signal gets intensified more than the legitimate signals.
As a result, the legitimate signals will be ``drowned out'' leading to increased packet retransmissions and reduced Quality-of-Service (QoS). We examine this phenomenon in detail to highlight the effect of the gain competition attack on the Signal-to-Noise Ratio (OSNR) and Bit Error Rate (BER) in an ONoC architecture. OSNR is a key parameter to estimate ONoC performance. It suggests a degree of impairment when the optical signal is carried by an optical transmission system that includes optical amplifiers. While optical amplifiers are essential to optical networks, they also have inherent vulnerabilities.
Amplifiers inherently degrade the OSNR as signals pass through them. The manufacturer specifies the amplifier's noise figure, which quantifies how much the signal degrades after amplification. This means that there will always be a lower OSNR after amplification.
The stronger signals receive more gain on amplification. This leads to a vulnerability called gain competition, where stronger signals receive more gain and can ``drown out'' the weaker signals~\cite{pasricha2020survey}.

Figure~\ref{fig:edfaBig} shows how the OSNR is affected in a gain competition attack. Consider initial OSNR ($OSNR_{0}$), which is an inherent property of the laser. 
At point $X$, we see the malicious member of the network injecting some strong signal on an out-of-band wavelength (represented by the red arrow). Note that the malicious signal may not be in the same band as the legitimate signal. An attacker can use a signal closer to a legitimate signal for maximum impact of the attack~\cite{furdek2010gain}. At the amplifier $Y$, the gain is stolen by the strong wavelength, and a relatively small gain is given to the legitimate signals (yellow, green, and blue wavelengths). The OSNR ($OSNR_{1}$) severely degrades compared to the original $OSNR_{0}$ after the amplification. This leads to an increased BER between points $Y$ and $Z$ and a drop in QoS~\cite{deng2002analysis}.

\subsection{Research Contributions}
There are various research efforts that examine different types of attacks that can occur on ONoCs~\cite{guo2020potential}~\cite{chittamuru2018soteria}. While the gain competition attack is well understood in the computer networks domain, it has not been explored in the context of ONoC design. This paper focuses on modeling the gain competition attack, exploring its impact on ONoCs, and proposing potentially effective mitigation techniques. Specifically, this paper makes the following contributions.

\begin{enumerate}
	\item We model gain competition attacks in optical NoCs using a cycle-accurate NoC simulator, Noxim~\cite{catania2015noxim}. We incorporate the bit error rate with the model to show the impact on system throughput and explore QoS and Denial-of-Service (DoS) attacks in ONoCs.
	\item We simulate the effect of the attacks on the performance of ONoCs and present our results to demonstrate the effectiveness of our modeling and exploration framework using a wide variety of traffic patterns and network topologies. We also discuss a few feasible mitigation techniques for the attacks in ONoCs.
\end{enumerate}

\section{Background and Related Work}
\label{section:background}
\subsection{Optical NoC (ONoC)}


\begin{figure}[t]
	\centering
\vspace{-0.2in}
	\includegraphics[scale=0.4]{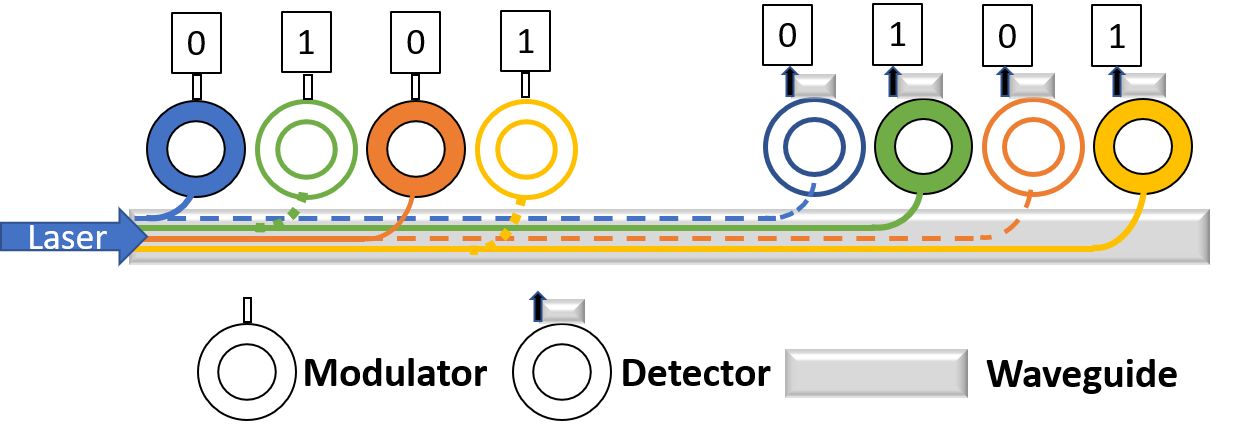}
 \vspace{-2mm}
	\caption{A DWDM-based photonic waveguide with modulators and detectors}
	\label{fig:MRdiagram}
\vspace{-0.2in}
\end{figure}

\begin{figure}[t]
	\vspace{-0.2in}
	\centering
	\includegraphics[scale=0.55]{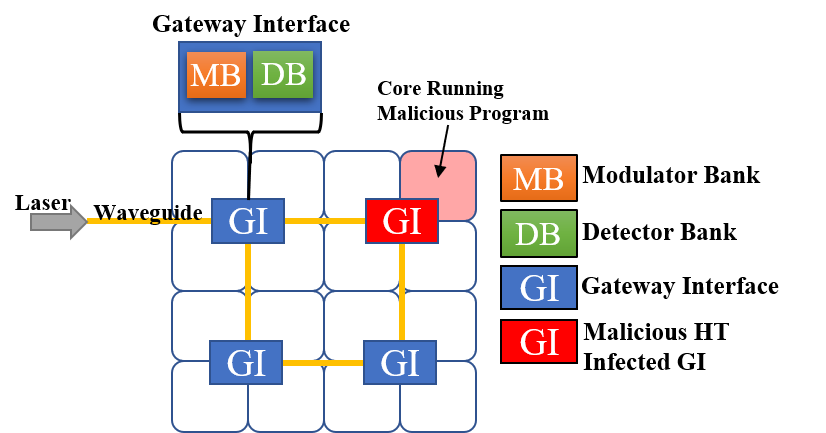}
    \vspace{-1mm}
	\caption{An instance of HT infected GI in 4 $\times$ 4 hybrid-NoC.}
	\label{fig:sudeepDiagram}
\vspace{-0.1in}
\end{figure}

Optical NoC provides the same set of benefits as electrical NoCs on large multi-core systems. 
In ONoC, the interconnects on the chip use light to send data instead of electrical signals, outperforming their electrical counterpart with near speed-of-light latency, increased bandwidth, and lower power dissipation~\cite{pasricha2020survey}~\cite{chittamuru2020exploiting}. ONoCs are often combined with electrical interconnects to form a hybrid topology (hybrid-NoC). This allows nearby cores to communicate electrically (like local streets) while optical communication is used (like a highway) for fast transmission over relatively long distances. These optical links are connected with an off-chip laser source.  
One of the key components of ONoC is Gateway Interface (GI). Each GI in an ONoC is able to send and receive data in the optical domain on multiple utilized carrier wavelengths. Therefore, each GI has two major components: the bank of MR modulators and a bank of detector MRs (DB), which are responsible for sending and receiving optical signals. In a bank, each MR operates on a specific carrier wavelength,  which allows the utilization of the excellent wavelength selectivity of MRs and DWDM capability of waveguides to enable high bandwidth parallel data transfers in ONoCs~\cite{pan2010flexishare}. The signals are sent along the waveguide to the destination GI, where the DB transforms the optical signals back to electrical ones. Figure~\ref{fig:MRdiagram} shows an example of a GI, with 4 MRs corresponding each to its wavelength used for DWDM communications~\cite{pasricha2020survey}~\cite{chittamuru2020exploiting}. GIs are also referred to as ``hubs'' later in the paper.





Figure~\ref{fig:sudeepDiagram} shows an illustrative example of a hybrid-NoC architecture with one core running a malicious program. The program triggers the infected GI, where the MR modulators and DB are found, and they become malicious in the network. The Trojan-infected GI gets activated under a specific condition, manipulating light and exploiting inherent vulnerabilities in the optical interconnect~\cite{chittamuru2020exploiting}. In following sections, we explore how these hardware Trojans can insert a strong jamming signal to commence a gain competition attack in ONoCs.

\subsection{Related Work}

There are several research efforts on exploring ONoC security vulnerabilities and developing effective countermeasures. Many of these vulnerabilities stem from possible Hardware Trojans (HT) in the network and the manipulation of MRs. Chittamuru et al.~\cite{chittamuru2018soteria} explored a situation where an HT is capable of manipulating the tuning circuit and can tune MRs to initiate a snooping attack. The paper proposes an encryption scheme where keys are based on the individual process variation profiles for each group of MR interfaces. 
Attacks that pose serious threats to the performance of electrical NoCs, such as \textit{blackhole} attacks, \textit{sinkhole} attacks, and \textit{flooding} attacks, were shown for ONoCs as well. 
Xie et al.~\cite{xie2010crosstalk} analyzed BER and OSNR of photonic data transmission in ONoCs. The crosstalk noise is generated when a small amount of the power from one signal interferes with other signals being used in the optical network. The noise leads to increased BER and limits the scalability of ONoCs due to degraded QoS. 
Xie et al. proposed a novel compact high-SNR optical router that improves the maximum size of mesh-base ONoC to 8$\times$8 network for BER of $10^{-3}$. Later in our paper, we utilize the range of BER values given for reliable transmission in~\cite{xie2010crosstalk} to model the proposed gain competition attack (see Section~\ref{section:gain_competition_model}). 
Other works also have shown the use of EDFAs~\cite{lee2009multi}~\cite{ji2011microring} and SOAs~\cite{thakkar2016run} in optical NoCs. The effect of introduced noise on OSNR drop and mitigation of the drop using the optimized placement of SOA in hybrid-NoC is shown in~\cite{jang2019power}.
Jang et al.~\cite{jang2019power} show the effect of introduced noise on SNR drop and how optimized placement of SOA can mitigate this problem in hybrid-NoC.

There are a few efforts to explore 
gain competition attacks in the computer networks domain~\cite{deng2002analysis}~\cite{xiaomin2015implementation}, the attack has not been explored in the context of optical on-chip communication. To the best of our knowledge, this paper makes the first attempt to investigate gain competition attacks as a security threat in ONoC-based SoCs.

\begin{figure}[t]
    \centering
        \vspace{-0.2in}
    	\includegraphics[scale=0.5]{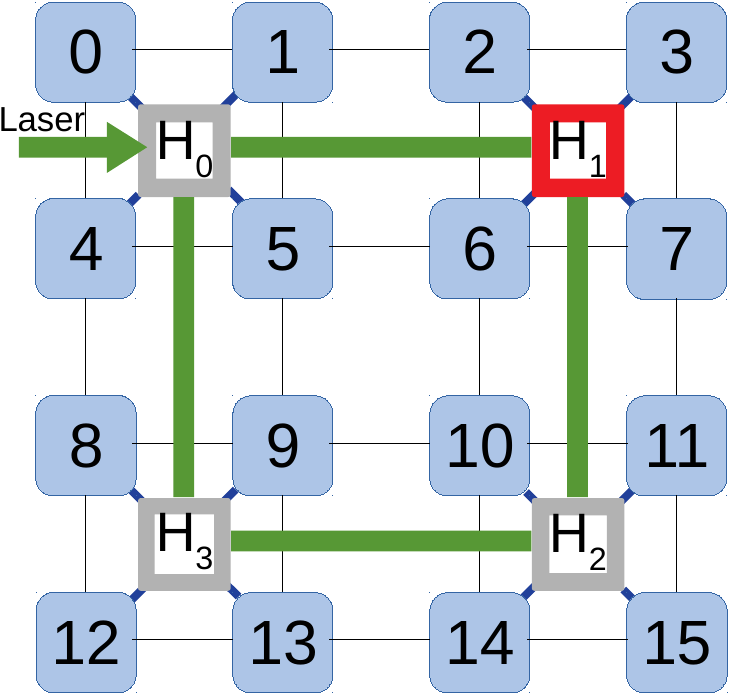}
    	\vspace{-2mm}
    	\caption{A 4$\times$4 hybrid-NoC with 4 optical hubs.}
    	\label{fig:16_core_example}
    	\vspace{-0.2in}
\end{figure}

\section{Gain Competition Attack}
\label{section:attack}



The gain competition attack utilizes the principle of gain competition that occurs when amplifying optical signals to deny or lower the quality of service in an optical network. In a gain competition attack, a malicious component such as HT-infected GI injects a powerful signal (usually 10 to 20dB stronger than other signals) of a wavelength different from the legitimate signals but still in the pass band of the amplifier. After the malicious signal is injected, it passes through an amplifier. The amplifier cannot distinguish between the attacking and legitimate signals. Therefore, it amplifies all the signal that passes through it. The attacking signal will receive more gain than the legitimate signals because it receives protons proportional to its power level~\cite{pasricha2020survey}. This will result in decreased OSNR and, consequently, increased BER. 

There are several factors, such as the power of the malicious signal, the wavelength distance with legitimate signal, channel spacing, the modulation speed, and the MR quality, on which gain competition attack depends. However, in our proposed attack model, we consider two important factors, the strength of the malicious signal and the distance of the maliciously injected signals from the legitimate signals, while inserting HT at the GI.
The closer the attacker's wavelength is to the legitimate wavelengths, the more detrimental the attack will be on the network. The same goes for the strength of the attacking signal. Note that a stronger attacking signal means more gain will be robbed at the amplifier, and the OSNR drop will be more severe~\cite{xiaomin2015implementation}~\cite{furdek2012physical}. 

The effectiveness of the gain competition attack increases when optical signals have to pass through multiple amplifiers, as the OSNR deteriorates at each amplifier. There are differences in each wavelength's available gain growth at each amplifier. If an attacker's injected signal goes undetected, this cascading amplifier effect can wreak havoc on an interconnection network. For larger ONoCs,  where the signal passes through multiple amplifiers, this can be a major problem \cite{furdek2012physical}.

\begin{figure}[t]
	\centering
	\vspace{-0.2in}
	\includegraphics[scale=0.4]{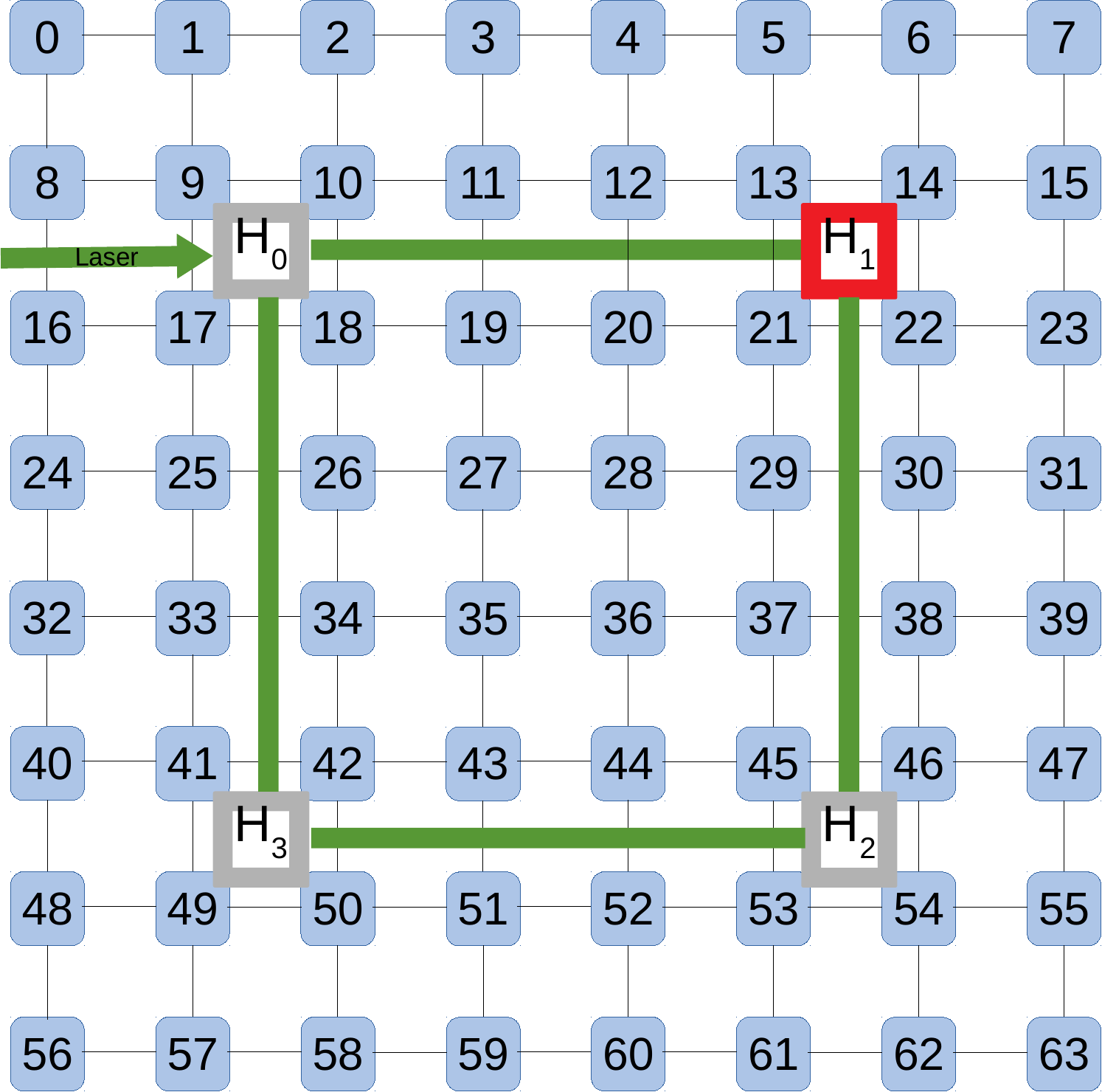}
	\vspace{-2mm}
	\caption{A 8$\times$8 hybrid-NoC with 4 optical hubs.}
	\label{fig:64_core_example}
 \vspace{-0.2in}
\end{figure}

\subsection{Modeling Gain Competition Attacks} 
\label{section:gain_competition_model}

We model a gain competition attack on hybrid-NoC using a widely-used Noxim~\cite{catania2015noxim}, cycle-accurate SystemC NoC Simulator.
For simplicity, we treat the wireless hubs in Noxim as optical hubs/GIs, assuming the performance to be comparable. In our model, a hub connects a cluster of cores to the photonic links necessary for optical data transmission~\cite{pasricha2020survey}~\cite{chittamuru2018soteria}. These hubs contain an array of MRs to send wavelength-multiplexed signals and an array of DBs to filter each wavelength and drop them on their respective photodetectors.

To emulate a gain competition attack, we assume one hub ($H_1$), as shown in Figure~\ref{fig:16_core_example} and~\ref{fig:64_core_example}, can maliciously inject wavelengths and decrease the QoS of a particular waveguide. IP cores that are connected to this hub can also initiate an attack if they are running some malicious software component. As shown in Figure~\ref{fig:16_core_example}, $H_1$ is under the control of an attacker and is able to inject wavelengths strong enough to cause gain competition. This means that the BER of the optical transmission from $H_1$ to $H_2$ would be significantly higher than $H_0$ to $H_1$. To model this, we modified Noxim and implemented a wireless BER parameter with the ability to apply it to specific hub-to-hub connections. 
The attacking wavelengths used by the attacker in $H_1$ would be the ones that are used for communication within the network but not for the particular instance of data transmission when the attack occurs. This makes the attack more covert and feasible than wavelengths unsupported by the on-chip optical components (i.e., laser, MRs, etc.). We see legitimate wavelengths used for gain competition attacks in~\cite{rejeb2006fault} as well.

\begin{figure*}[t]
	\begin{minipage}[]{\columnwidth}
        \vspace{-0.1in}
		\centering
		\includegraphics[width=1\columnwidth]{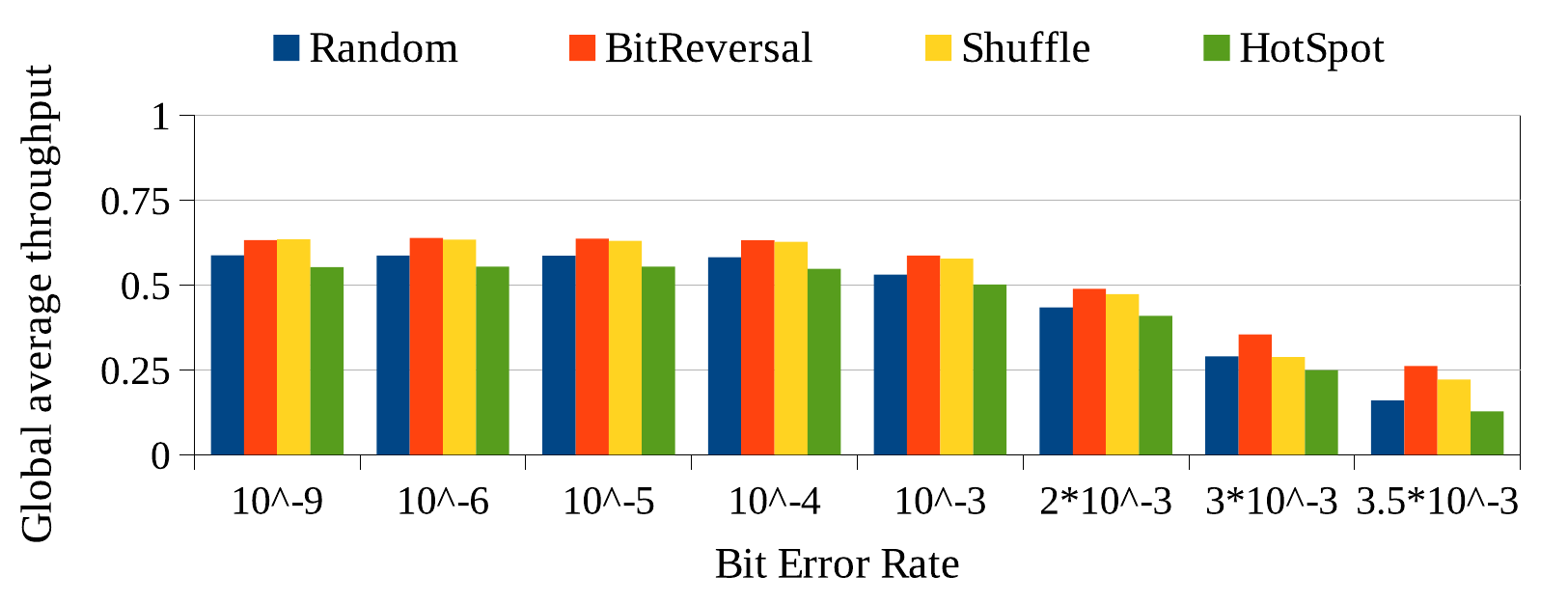} 
	    \vspace{-0.3in}
	    \caption{GAT (flits/cycle) for different BER values in 4$\times$4 hybrid-NoC}
        \label{fig:global_average_throughput}
    \end{minipage}
\hfill
    \begin{minipage}[]{\columnwidth}
        \vspace{-0.1in}
    	\centering
    	\includegraphics[width=1\columnwidth]{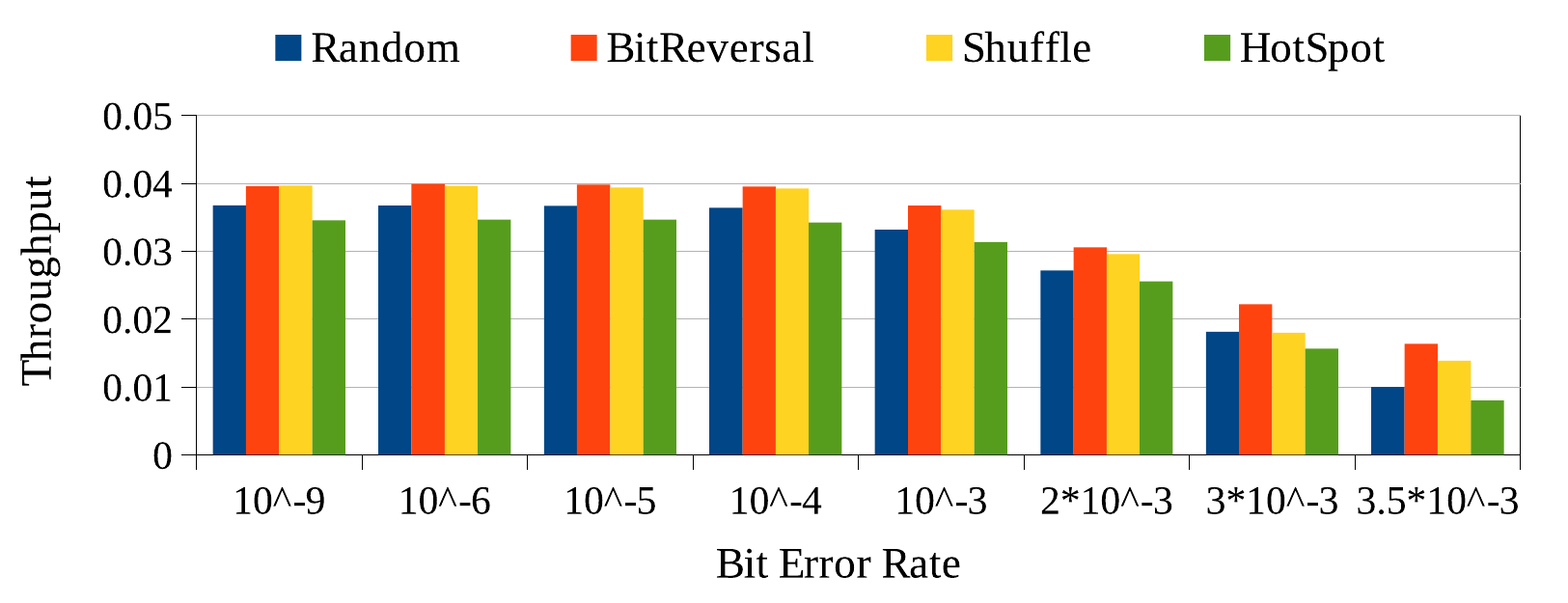}
    	\vspace{-0.3in}
    	\caption{Throughput for different BER values in 4$\times$4 hybrid-NoC}
    	\label{fig:throughput_ip}
    \end{minipage}
    \vspace{-0.2in}
\end{figure*}

We set the wireless BER to $10^{-3}$ to model a hub-to-hub connection that has undergone a gain competition attack. This means that $1$ in every $1000$ bits is erroneous, causing any packet containing an erroneous bit to be retransmitted.
We adapt the range of BER values given for reliable transmission in~\cite{xie2010crosstalk} for our proposed attack modeling. For reliable optical transmission, the BER has to be in the range of $10^{-9}$ to $10^{-3}$. Our model uses packet retransmission probability to analyze the effect of the gain competition attack on NoC.
The packet retransmission probability ($P_{retransmit}$) can be calculated as $BER$ $\times$ $f$ $\times$ $p$, where $f$ is flit size and $p$ is packet size in flits.
For example, with a BER of $10^{-3}$, a flit size of $32$ bits, and a packet size of $8$ flits,  $P_{retransmit}$ $=$ $10^{-3}$ $\times$ $32$ $\times$ $8$ $=$ $0.256$.
Roughly one out of four packets must be retransmitted on the compromised optical connection. When packet retransmission is necessary, we model a delay equivalent to one round trip period to account for the transmission delay of the failed acknowledgment back to the source and retransmit the packet.

Our framework can model gain competition attacks on various NoC configurations and network topologies executing diverse traffic patterns as discussed in Section~\ref{section:experiment}. For example, Figure~\ref{fig:64_core_example} shows the placement of hubs in an $8$$\times$$8$ mesh hybrid-NoC with $64$ cores. The number of optical hubs stayed the same as the $4$$\times$$4$ network in Figure~\ref{fig:16_core_example}, but they are placed in the center of each of the four $16-core$ quadrants, where each core is at most three hops away from the optical hub. Note that such a hybrid model introduces interesting routing considerations. For example, a communication between the $Node_1$ and $Node_{55}$ can utilize the path: $Node_1 \rightarrow Node_9 \rightarrow H_0 \rightarrow H_2 \rightarrow Node_{54} \rightarrow Node_{55}$. However, a communication between the $Node_1$ and  $Node_6$ have many alternatives including only electrical ($Node_1 \rightarrow Node_2 \rightarrow Node_3 \rightarrow Node_4 \rightarrow Node_5 \rightarrow Node_6$) and hybrid ($Node_1 \rightarrow Node_9 \rightarrow H_0 \rightarrow H_1 \rightarrow Node_{14} \rightarrow Node_6$). Our modeling framework will enable designers to explore different routing protocols with varying optical hub placements under conflicting design constraints, including security, energy efficiency, and QoS.

\section{Experimental Evaluation}
\label{section:experiment}

\begin{figure*}[t]
\vspace{-2mm}
\begin{minipage}[]{\columnwidth}
	\centering
	\includegraphics[width=1\columnwidth]{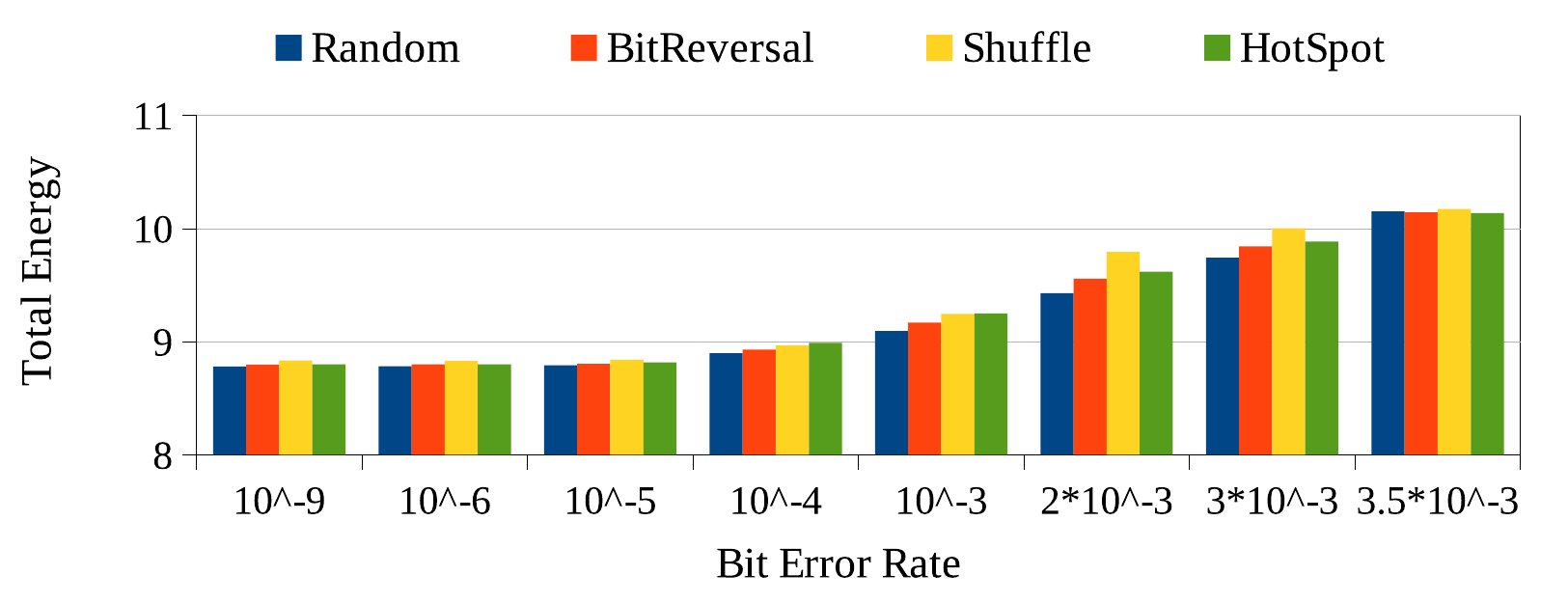}
	\vspace{-0.3in}
	\caption{Total energy usage ($\mu J$) for different BER values in 4$\times$4 hybrid-NoC}
	\label{fig:energy}
\end{minipage}
\hfill
\begin{minipage}[]{\columnwidth}
	\centering
	\includegraphics[width=1\columnwidth]{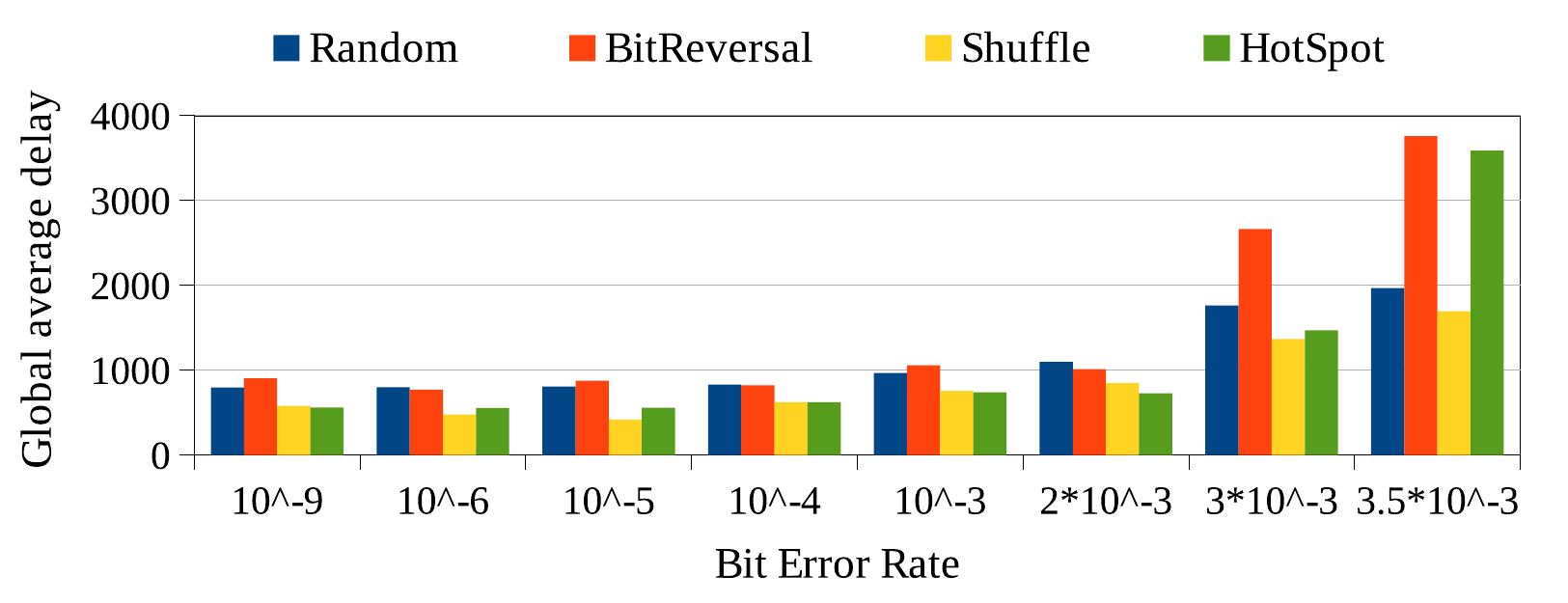}
	\vspace{-0.3in}
	\caption{GAD (cycles) for different BER values in 4$\times$4 hybrid-NoC}
	\label{fig:global_average_delay}
\end{minipage}
\vspace{-0.2in}
\end{figure*}

\begin{figure}[h]
\begin{minipage}[]{\columnwidth}
\centering
\includegraphics[width=1\columnwidth]{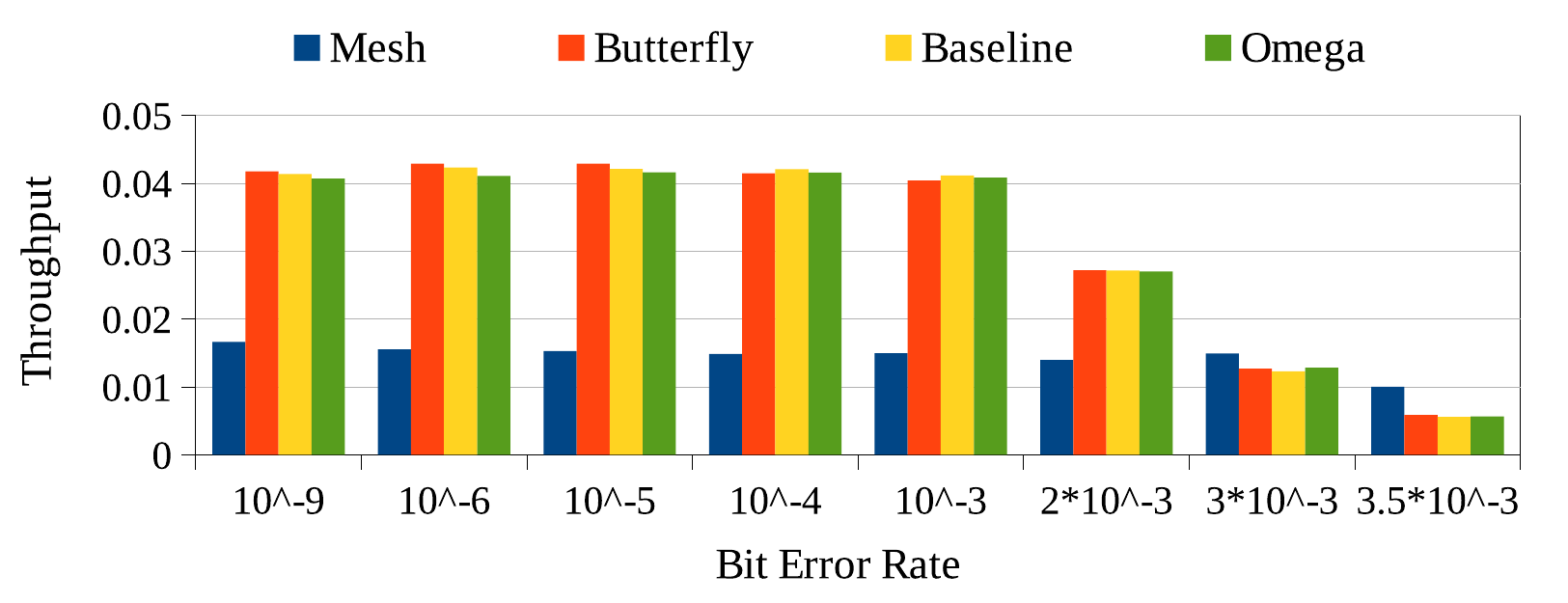}
\vspace{-0.3in}
\caption{Throughput (flits/cycle/IP) for different BER values for four different topologies in 4$\times$4 hybrid-NoC.}
\label{fig:throughput_ip_topology}
\end{minipage}
\hfill
\hfill
\begin{minipage}[]{\columnwidth}
\centering
\includegraphics[width=1\columnwidth]{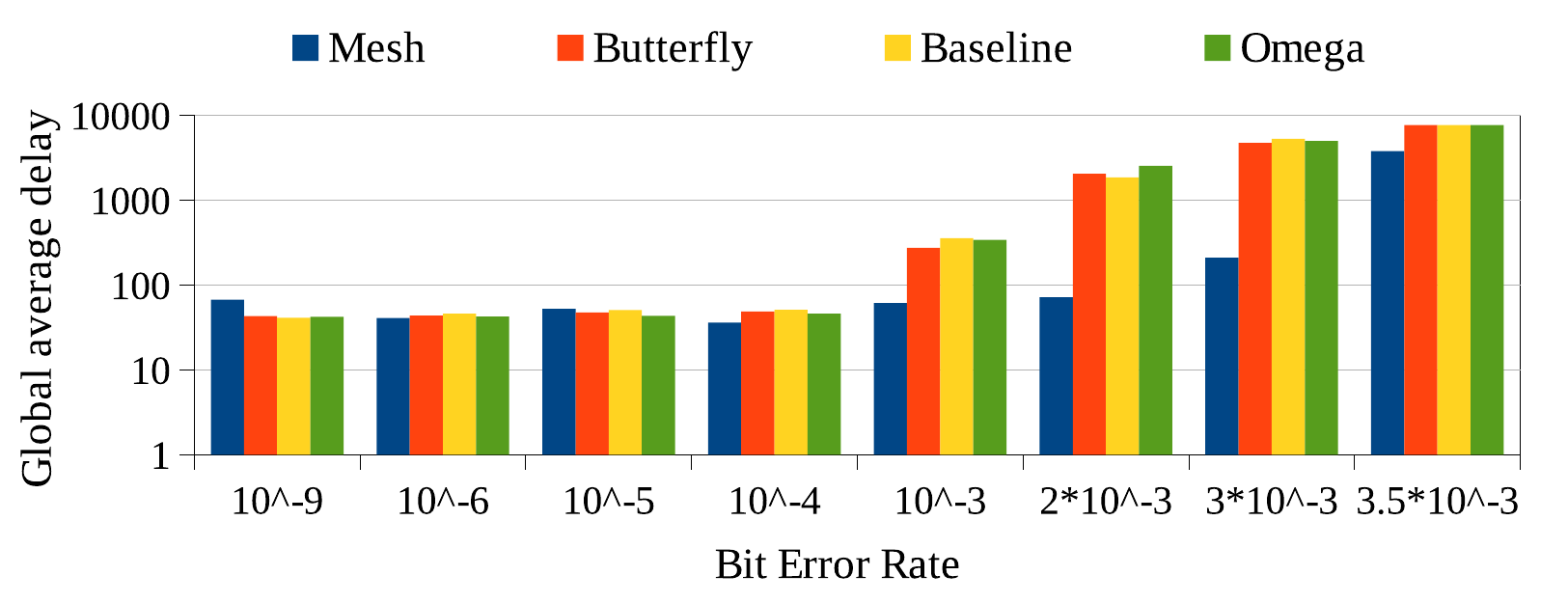}
\vspace{-0.3in}
\caption{GAD (cycles) for different BER values for four different topologies in 4$\times$4 hybrid-NoC.}
\label{fig:global_average_delay_topology}
\end{minipage}
\vspace{-0.2in}
\end{figure}

We demonstrate the applicability and scalability of our proposed attack on Noxim Simulator~\cite{catania2015noxim}, as described in Section~\ref{section:gain_competition_model}. We evaluate the attack on a 4$\times$4 and an 8$\times$8 hybrid-NoC with $4$ optical hubs as shown in Figure~\ref{fig:16_core_example} and Figure~\ref{fig:64_core_example}, respectively. 
In the network, all the routers are divided into 4 groups. Within each group, the routers are electrically connected using mesh topology, and each router in the group is connected to a different group through an optical link. 
We simulated the model using four synthetic traffic patterns {\it Random}, {\it BitReversal}, {\it Shuffle}, and random with a {\it HotSpot} on $Node_3$, with a fixed injection rate of 0.005 flits/cycle/node. 
We adopt the bit-error rate given in~\cite{xie2010crosstalk} to study the impact of the proposed attack for different BER values in the range of $10^{-9}$ to $3.5\times10^{-3}$.
We have also shown the behavior of the attack on a variety of network topologies such as {\it Mesh}, {\it Butterfly}, {\it Baseline~\cite{lu2004parallel}}, and {\it Omega}. 
Note that we have used hybrid-NoC to evaluate the proposed attack. However, the attack is equally applicable to Optical NoC.

For simplicity, in our experiment, we demonstrate the effect of the attack on one wavelength, as opposed to a wavelength-division-multiplexed network, where multiple wavelengths are used for data transmission.
Note that gain competition attacks can affect many wavelengths simultaneously and cause DoS attack. The magnitude of the effect on other wavelengths depends on the strength of the attacking signal and the difference in wavelengths between the attacker and the legitimate signals~\cite{furdek2012physical}.

\subsection{Results using 4$\times$4 Mesh hybrid-NoC}
\label{section:4x4}

In this section, we analyze the impact of gain competition attacks on throughput, energy consumption, and Global Average Delay (GAD).
Figure~\ref{fig:global_average_throughput} shows the effect of the  attack on Global Average Throughput (GAT). GAT represents the successful transmission of flits-per-cycle at the destination nodes. As shown in the graph, performance decrease results from the packet drops due to the compromised optical link with an elevated BER. This leads to packet retransmission and a decrease in network throughput.
Across all types of traffic, the network throughput decreased by an average of $68$\%, when comparing the least BER ($10^{-9}$) to the maximum BER ($3.5 \times10^{-3}$) in our evaluation. 
Among all four synthetic traffic patterns, \textit{BitReversal} showed the least throughput degradation ($58$\%), whereas \textit{HotSpot} suffered from maximum decrease ($77$\%) due to traffic hotspot generated at $Node_3$. $Node_3$ resides in the corner and far away from other nodes, making it a prime candidate for optical transmission (c.f. Figure~\ref{fig:16_core_example}). 
In traffic patterns, \textit{BitReversal} and \textit{Shuffle}, the number of packets traversing through the malicious link is significantly low compared to other traffic patterns. Unlike \textit{HotSpot}, this reduces the cases of packet retransmission. 
Figure~\ref{fig:throughput_ip} shows the average throughput per core in flits/cycle/IP. It shows a trend similar to that of Figure~\ref{fig:global_average_throughput} except being on a per-core scale, demonstrating that individual cores also experience degraded performance.

Figure~\ref{fig:energy} shows the effect of the gain competition attack on the NoC's energy consumption. As the BER increases, the energy of the network consumption also increases ($15$\%). This can be attributed to the energy required to retransmit erroneous packets caused by the gain competition attack. We use the default energy settings in Noxim to generate these results.

Figure~\ref{fig:global_average_delay} shows the global average delay of flits in the network. The delay is expected to be higher during a gain competition attack. Flits transmitted on compromised waveguides are more likely to have to be retransmitted, causing other flits to be delayed.
One outlier in this data set is the \textit{HotSpot} traffic pattern in the worst BER scenario. The \textit{HotSpot} traffic pattern makes $Node_3$ a very highly trafficked node for higher BER values, and therefore, the global average delay is $6.4$ times higher in BER $3.5\times10^{-3}$ over $10^{-9}$. The optical waveguides sending data to and around $Node_3$ (c.f. Figure~\ref{fig:16_core_example}) become highly congested in the most extreme attack scenario. With such a high BER, packets get retransmitted frequently, and the whole network becomes congested, as flits must wait to be delivered for an alarmingly long time.

\subsection{Analyzing the effect on different network topologies}

\begin{figure*}[ht]
\vspace{-0.2in}
\begin{minipage}[]{\columnwidth}
\centering
\includegraphics[width=1\columnwidth]{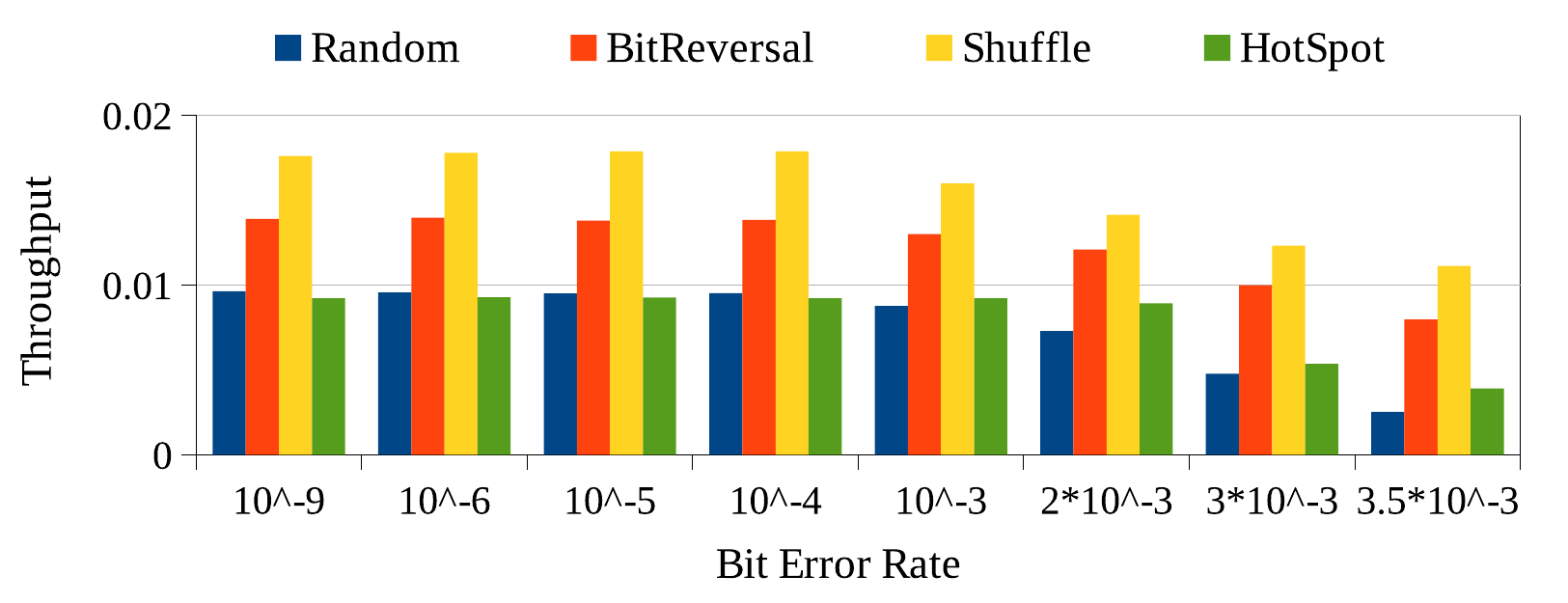}
\vspace{-0.2in}
\caption{Throughput for different BER values in 8$\times$8 hybrid-NoC.}
\label{fig:throughput_ip_64}
\end{minipage}
\hfill
\begin{minipage}[]{\columnwidth}
\centering
\includegraphics[width=1\columnwidth]{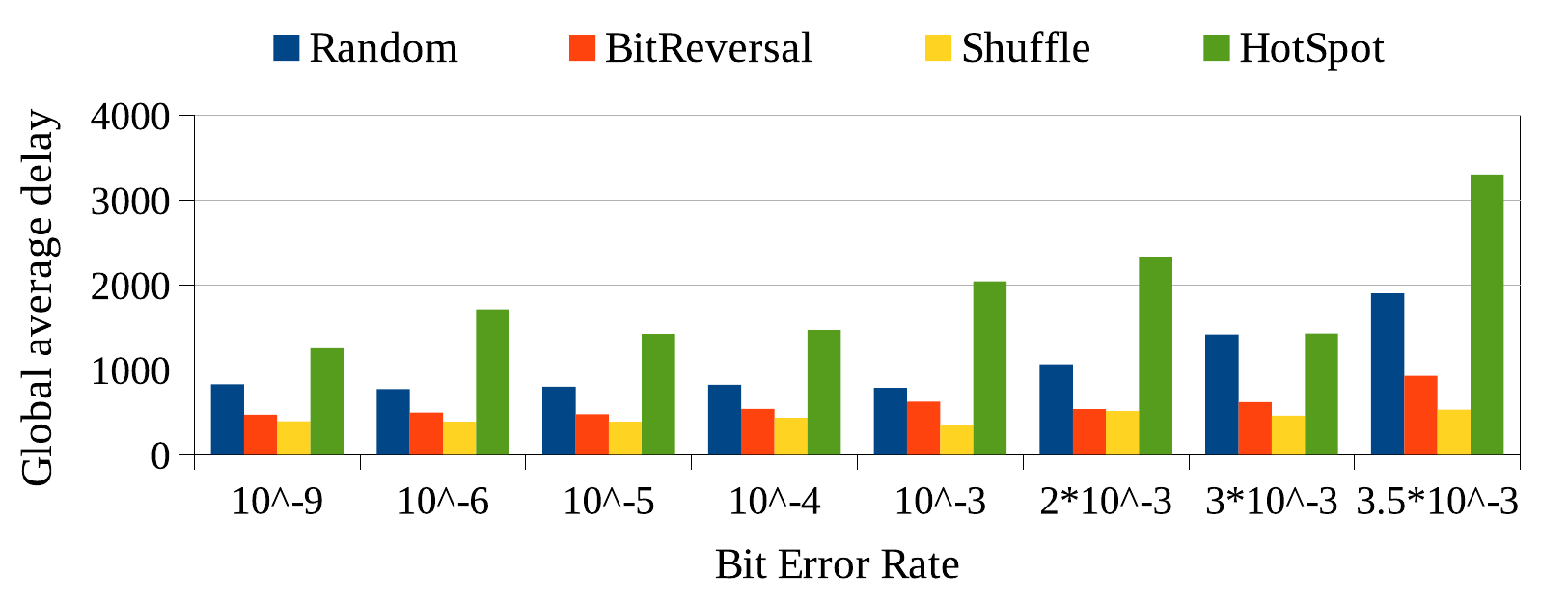}
\vspace{-0.2in}
\caption{GAD (cycles) for different BER values in 8$\times$8 hybrid-NoC.}
\label{fig:global_average_delay_64}
\end{minipage}
\vspace{-0.2in}
\end{figure*}

The previous section explored different traffic patterns on a 4$\times$4 mesh topology. Here, we explore the effect of gain competition attacks on the average throughput and delay for different network topologies. Figure~\ref{fig:global_average_delay_topology} shows the global average delay on {\it Mesh}, {\it Butterfly}, {\it Baseline~\cite{lu2004parallel}} and {\it Omega} topology with \textit{Random} traffic pattern. As we can see, a trend similar to Figure~\ref{fig:global_average_delay}, where the global average delay is higher for the higher bit error rates, is found on the right side of the graph. Similar behavior is observed for average throughput on four topologies (shown in Figure~\ref{fig:throughput_ip_topology}). This demonstrates that our model is applicable across different NoC topologies.

\subsection{Analyzing the effect on a bigger network}
\label{section:8x8}

As shown in Figure~\ref{fig:throughput_ip_64} and Figure~\ref{fig:global_average_delay_64}, we examine the average throughput per core and average delay for 8$\times$8 mesh topology (c.f. Figure~\ref{fig:64_core_example}). The results show the trend of network throughput decreasing as BER increases. The resulting difference for each traffic pattern is more pronounced on the 8$\times$8 setup than on the 4$\times$4. As expected, the throughput is close to the case when there is no attack and the BER value is zero. The global average delay (GAD) increases with the BER value. Similar to the 4$\times$4 setup, GAD is too high for \textit{HotSpot} traffic for all BER values. In the bigger network, the congestion at $Node_3$ is much higher even at the low BER values. This demonstrates that our gain competition attack modeling is applicable across different NoC configurations and traffic patterns.

\section{Attack Detection and Mitigation}
\label{section:mitigation}

In this section, we discuss a few potential gain competition attack detection and mitigation techniques. 
\textbf{Attack Detection:}
Attack detection involves three steps: recognition, classification and localization, and alarm generation~\cite{medard1997security}.
We explore the suitability of attack detection from the computer networks domain in ONoC~\cite{furdek2012physical}. 
Optical Spectrum Analyzer (OSA) provides a more detailed diagnosis than typical wideband power detection. It can detect a change in the shape of the optical spectrum used for optical data transfer. 
OSA measures the individual wavelengths and amplitudes and generates an alarm on attack detection. Optical spectrum analysis is a statistical method and requires averaging over time, making it slower than other methods and potentially unable to detect infrequent degradation of the signal~\cite{furdek2012physical}~\cite{medard1997security}. 
In~\cite{li2015integrated}, experiments were done with an on-chip OSA device that effectively measured the amplitude of light waves respective to their wavelengths. This makes OSA a strong option for defending against gain competition attacks in ONoCs.

Another statistical method discussed in~\cite{saengudomlert1998analysis} measures the BER of optical transmissions over a given period using a piece of hardware called a $BER-Tester$. This method effectively detects attacks when the quality of an optical transmission line is degrading due to elevated BER. 
However, it requires averaging and comparison, which can delay attack detection. 
Monitoring the dropped/retransmitted packets in the network and comparing them to the threshold can eliminate the hardware overhead involved with a $BER-Tester$ in ONoC. 


Pilot tones are another way to detect the occurrence of a gain competition attack. They are often out-of-band signals that travel along the same paths as communication signals but are distinguishable. When it passes through the same EDFA as the legitimate communication signals, the pilot tones' quality will be degraded, and the attack can be detected by monitoring the degradation. This method may fail if the OSNR is not degraded enough to reach the detection threshold~\cite{furdek2012physical}~\cite{medard1997security}. 



\textbf{Potential Countermeasures:}
Once an attack is detected, several responses can happen. The direct response to a gain competition attack is to find and eliminate the attacking wavelength. OSA can be used to detect the attack, and measures can be taken to eliminate the wavelength interfering with legitimate signals. If the above response is not feasible, traffic can be rerouted via unaffected parts of the on-chip network. In our example (Figure~\ref{fig:16_core_example}), there is an option to send traffic along the electrical wires in the network. Sending packets through the electrical route may cause congestion and be slower; however, this may be a good option if the BER is too high in the optical waveguide.

Another way to route packets around a gain competition attack is to send packets on wavelengths that are far away from the jamming signal's wavelength. Several experiments have shown that the BER is lower for data transmission on wavelengths further away from the attacker's wavelength in a gain competition attack scenario~\cite{xiaomin2015implementation}~\cite{furdek2012physical}. 
This requires a transmission spectrum large enough to contain wavelengths that are adequately far from the attacker's wavelength to have a BER acceptable for optical transmission. 

The variant of the $BER-Tester$ approach that uses thresholds appears to be the best fit for our threat model, considering the resource-constrained nature of ONoC. However, it is worth noting that the most beneficial detection method and actions can depend on the architecture and threat models for other instances.

\section{Conclusion}
\label{section:conclusion}

Optical network-on-chip architectures provide various advantages over its electrical counterpart. However, optical NoCs face various security challenges. In this paper, we investigated gain competition attacks in optical NoCs. Specifically, we developed a method to model a gain competition attack using optical bit error rate. We explored the impact of a gain competition attack using several optical NoC configurations, traffic patterns and topologies.  Our experiments using different NoC topologies and diverse traffic patterns demonstrated that gain competition attacks present a serious threat to optical NoCs. We proposed several promising attack detection as well as attack mitigation techniques that would enable design of trustworthy optical NoCs.


\bibliographystyle{IEEEtran}
\bibliography{IEEEabrv,bibliography}

\end{document}